# Triggered Ion-acoustic Waves

By Forrest Mozer[1], Stuart Bale[1], Paul.Kellogg[3], Orlando Romeo[2], Ivan Vasko[2], and Jaye Verniero[4]

1. Physics Department and Space Sciences Laboratory, University of California, Berkeley, 94720
2. Space Sciences Laboratory, University of California, Berkeley, 94720
3. University of Minnesota, Minneapolis, Minnesota, 55455
4. Goddard Space Flight Center, Greenbelt, Md., 20706

Corresponding author: Forrest.Mozer@gmail.com

Triggered ion-acoustic waves are a pair of coupled waves observed in the previously unexplored plasma regime near the Sun [Mozer et al, 2021; 2022a]. They may be capable of producing important effects on the solar wind. Because this wave mode has not been observed or studied previously and it is not fully understood, the issue of whether it has a natural origin or is an instrumental artifact can be raised. This paper discusses this issue by examining 13 features of the data such as whether the triggered ion-acoustic waves are electrostatic, whether they are both narrow-band, whether they satisfy the requirement that the electric field is parallel to the k-vector, whether the phase difference between the electric field and the density fluctuations is 90 degrees, whether the two waves have the same phase velocity as they must if they are coupled, whether the phase velocity is that of an ion-acoustic wave, whether they are associated with other parameters such as electron heating, whether the electric field instrument otherwise performed as expected, etc. The conclusion reached from these analyses is that triggered ion-acoustic waves are highly likely to have a natural origin although the possibility that they are artifacts unrelated to processes occurring in the natural plasma cannot be eliminated. This inability to absolutely rule out artifacts as the source of a measured result is a characteristic of all measurements.

I  SUMMARY OF TRIGGERED ION-ACOUSTIC WAVE PROPERTIES

The Parker Solar Probe has measured triggered ion-acoustic waves (TIAW) in each of the seven orbits that passed through the altitude range of 15-25 solar radii [Mozer et al, 2021; 2022a]. They have the following properties:
- A pair of coupled waves, one of which is at a frequency of a few Hertz and the other is at few hundred Hertz
- The two waves are coupled such that the high frequency wave sometimes occurs in short bursts during successive periods of the low frequency wave and at a fixed phase of this lower frequency wave.
- The waves are electrostatic.
- Both the high and the low frequency waves are narrow band, effectively pure sine waves.
- They can exist as a pair for times as long as several hours.
- They are associated with heating the core electron distribution.

The Parker Solar Probe is in a solar orbit with one surface (the heat shield) facing the Sun near perihelion. The line of sight from the spacecraft to the Sun is the spacecraft Z-direction, along which the typical magnetic field and solar wind flow are oriented near perihelion. The X-Y plane, perpendicular to the Sun-satellite line, contains a two-component electric field and spacecraft potential measurement by antennas that are not much larger than the spacecraft [Bale et al, 2016]. By fitting the measured spacecraft potential to the low-rate density measurements obtained from the SWEAP plasma measurements [Kasper et al, 2016; Whittlesey et al, 2020], higher frequency estimates of the plasma density and density fluctuations are obtained [Mozer et al, 2022b]. In addition, the SWEAP/SPAN-I [Livi et al, 2022] instrument provided the solar wind velocity utilized in the following analyses. The solar wind speed, magnetic field, plasma density, ion temperature and other parameters for the event that follows are given in Table 1. All data presented in this paper are in the spacecraft coordinates whose X, Y, and Z directions are defined above.

Figure 1 presents TIAW observed on January 19, 2021. The four pairs of panels each display EX and the density fluctuations, $\delta n/n$. Panel (1a), gives 1 Hz high pass filtered electric field data that illustrates the low and high frequency waves that are phase locked such that the high frequency bursts occur at fixed phases of the low frequency signal. Panel (1b) illustrates the density fluctuations measured above 1 Hz and they show mainly the low frequency wave signal. Panels (1c) and (1d) show the low frequency (0.5-2 Hz) components of the TIAW with $\delta E$~0.5 mV/m and $\delta n/n$ ~0.13, while panels (1e) and (1f) illustrate the high frequency (>100 Hz) components having $\delta E$~4 mV/m and $\delta n/n$ ~0.004. Panels (g) and (h) present a short portion of the high frequency components to illustrate that they are nearly pure sine waves, which is not expected for ion-acoustic waves (the full spectral width at half-maximum is less than 60 Hz). Figure 1 presents the same event as that described in Figure 6 of Mozer et al [2021] and it was selected to benefit from this earlier information that shows the phase locking of the low and high frequency waves and the absence of a magnetic field signature during such events. It is important to note that the potentials of antennas 1 through 4 produced both the electric field and the density data because the difference between potential pairs is the electric field and the sum of the four potentials is the spacecraft potential that gives the plasma density [Mozer et al, 2022b]. Thus, although the origins of the electric field and density are related, they are really two different quantities.

II ARGUMENTS CONCERNING WHETHER TRIGGERED ION-ACOUSTIC WAVES ARE A NATURAL PHYSICAL PHENOMENON OR AN INSTRUMENTAL ARTIFACT.

**IIa The TIAW occur at times of electron heating in the solar wind. Most importantly, in the absence of TIAW, electrons are not heated at 20-25 solar radii.**

Triggered ion-acoustic waves are seen at 20-25 solar radii on practically every orbit where they have been searched for. Mozer et al [2022a] showed this to be true for orbits 6, 7, 8, and 9, and more recent observations on orbits 10, 11, and 12 also confirm their presence. The core electron temperature has been observed to increase in conjunction with all triggered ion-acoustic wave events [Mozer et al, 2022a] and, most importantly, such heating is absent when there are no triggered waves [see Figure 1 of Mozer et al, 2022a]. This correlation may exist because core electrons are heated by the triggered ion-acoustic waves or, conversely, because the increased electron temperature is favorable for generation of TIAWs.

Other non-wave electron heating mechanisms may also operate in the solar wind. For example, Boldyrev et al [2020] proposed that, due to conservation of their magnetic moments, solar wind electrons form a beam collimated along the magnetic-field lines. Due to weak energy exchange with the background plasma, the beam population slowly loses its energy and heats the background plasma. In this model, electrons are heated more or less uniformly as a function of distance while the observed heating largely occurs between 20 and 25 solar radii [Mozer et al, 2022a] and only when TIAW are present.

The amplitudes, $\delta\varphi$, of the electrostatic potential of both TIAW waves can be estimated in the following way. Because the observed waves are expected to have phase speeds much smaller than the electron thermal speed, the parallel electric field should be balanced by the electron pressure gradient [Davidson, 1972; Kelley and Mozer, 1972]

$$enE_\parallel = -\nabla_\parallel p_e \qquad (1)$$

where n is the plasma density, and $\nabla_\parallel p_e$ is the parallel electron pressure gradient, $\nabla_\parallel nkT_e$. Thus, $\delta\varphi/T_e \approx \delta n/n$. With the local electron temperature for the interval in Figure 1 of about 54 eV (Table 1), we find for the high and low frequency electrostatic waves that

$$\delta\varphi_h \approx 0.2\ V \quad \text{and} \quad \delta\varphi_l \approx 7\ V \qquad (2)$$

Since ion-acoustic waves have speeds much slower than the electron thermal speed, the electron heating due to their interaction with these waves via the Landau resonance can only be less than or the order of their potentials. Thus, the direct electron heating due to the high-frequency waves cannot exceed 1%, while the direct heating due to the low-frequency waves can be as large as 10%. This amount of heating is generally consistent with the core electron temperature increases that have been observed at the times of these waves [Mozer et al, 2022a].

**IIb The high and the low frequency waves are both electrostatic.**

Panels (b) and (c) of Figure 1 of Mozer et al, [2021] show prominent signatures in the electric field spectra during a 20-hour interval containing the high frequency component of the triggered ion-acoustic wave that varied from 1000 to 200 Hz. Panels (e) and (f) of this figure cover the same time and frequency interval for two components of the magnetic field measured by the search coil magnetometer. Because there is no corresponding signal in the magnetic field data, there is no magnetic field in the high frequency wave. For a 100 km/sec phase velocity of these waves [Mozer et al, 2021], the magnetic field associated with a phase speed E/B~100 km/sec in the presence of the ~1 mV/m low frequency electric field of Figure 1a, if it was an electromagnetic wave, would be ~10 nT. The upper limit to the observed magnetic field at this low wave frequency is at least an order of magnitude less, showing that the low frequency wave also does not have a magnetic component. Thus, the waves of interest must be electrostatic. These results are expected from Figure 1 which shows correlated electric field and density fluctuations in both the low and high frequency waves

**IIc The electric field fluctuations in the high frequency wave are parallel to the k-vector in the X-Y plane, as they must be if the wave is real.**

Figure 2 provides detailed information on the event of Figure 1 that enables a measurement of the angle between the electric field and the k-vector in the X-Y plane. As seen in panel (2a), the 600 Hz electric field oscillated in the ±X direction while the component of the solar wind velocity in the X-Y plane was in the –X direction (panel 2c) and the magnetic field in the X-Y plane was inclined about 30º degrees away from the +X-axis (panel 2b). The direction of wave propagation may be determined from timing the crossing of the wave over each of the antennas. Panel (2d) shows that antennas V1 and V4 received the maximum wave signal at about the same time. From the geometry illustrated in Figure 3, this requires that the wave traveled in either the +X or –X direction. If it traveled in the –X direction, the signals on V1 and V4 would precede the signals on V2 and V3, as actually happened in panel (2d). (The negatives of the signals on V2 and V3 are plotted to compensate for the fact that the peak of a wave traveling to the left produces negative voltages on V2 and V3 because they measure the voltage on the antenna minus the voltage of the spacecraft body). In summary, in the spacecraft X-Y plane, the wave propagated in the –X direction and the electric field oscillations were along this direction. Thus, in the X-Y plane the requirement that E must be parallel to k is satisfied.

**IId The phase velocity of the high frequency wave is consistent with that of an ion-acoustic wave**

The high frequency wave speed in the X-Y plane may be determined from the time lags of the single ended potentials in Figure (2d). The time difference between the wave reaching V1/V4 and V2/V3 was about 0.53 msec and the wave traveled about three meters during this time. Thus, in the spacecraft X-Y frame, the wave speed in the X direction was -6 km/sec. Because the solar wind speed in the X direction was -90 km/sec in panel (2c), the wave speed in the plasma frame was in the X direction at about -84 km/sec. Neither the electric field nor the wave speed in the Z direction were measured so the total phase velocity cannot be determined. However, from the measured wave speed in the X-Y plane, a total phase velocity that is consistent with an electrostatic wave seems quite plausible.

**IIe To be physical, the two TIAW must have the same phase velocity because they are coupled. Direct measurements of the phase velocities in the plasma frame show that they are equal in magnitude and direction in the X-Y plane to well within experimental uncertainties.**

It is complicated to measure the phase velocity of the low frequency wave because, at its frequency of 1.25 Hz, there are also electromagnetic fields due to Alfvenic turbulence, as illustrated in Figure 4. In panel (4a), the x-component of the filtered low frequency electric field is plotted. It is a nearly pure sinusoid, which signifies that the electrostatic field was much greater than the electromagnetic field for this component. By comparison, EY in panel (4b) deviates significantly from a sine wave. That this is because of the predominance of the y-component of the electromagnetic wave is seen by comparing EY with the similar wave form of BX in panel (4c). Thus, it is concluded that the low frequency electrostatic wave oscillated primarily in the x-direction.

Figure 5 presents the band pass filtered low frequency potentials on the four electric field antennas in panels (5a) through (5d). That the amplitudes of the signals and, to some extent, their frequencies vary within the bandwidth is due to the electromagnetic field contamination. Determining the time differences between these various voltage signals to obtain the low frequency phase velocity is less accurate for this reason as well as, for the reason that because of the low frequency of the wave, a short time difference is more difficult to determine than is the case for the high frequency wave of section IId. Nevertheless, the cross correlation of antenna pairs as a function of the lag time is presented in panel (5e) in order to obtain approximate information on the low frequency wave propagation direction and speed. That the time lags between signals on V2 and V3 (the red curve) as well as between signals on V1 and V4 (the green curve) are minimum suggests that the wave propagated in $\pm X$ direction, as did the high frequency wave. That the correlation between V1 and V2 (the black curve) is maximum for -2 milliseconds suggests both that the wave propagated in the -X direction and that it propagated at a speed of about 2 km/sec. Thus, in the plasma frame, the low frequency wave traveled in the X-direction at about -88 km/sec. This velocity is consistent in magnitude and direction with the -84 km/sec speed of the high frequency wave. And thus, the low and high frequency waves traveled in the X-Y plane in the same direction and at the same phase velocity in the plasma frame, as they must in order to be coupled. It is noted that the near equality of the two phase speeds results from the fact that their phase speeds in the spacecraft frame were small, such that the exact values of the phase speeds in the spacecraft frame does not matter.

**IIf The phase difference between the electric field and density fluctuations in the high frequency wave was 90 degrees, as must be the case for real electrostatic waves.**

Because of equation (1), the electric field and the density fluctuations must be 90 degrees out of phase. As seen in Figure 1 panel (1g) and panel (1h), this requirement is fulfilled by the high frequency wave of the triggered ion-acoustic wave pair. Because of the presence of additional electric fields in electromagnetic waves at low frequencies, this requirement is hard to verify for the low frequency electrostatic wave although it is suggested by comparison of the plots in Figures (1c) and (1d).

**IIg The TIAW are observed to persist for up to several hours at a time. This result may be understood theoretically.**

Normal ion-acoustic waves have shorter durations than do the TIAW [Kurth et al., 1979; Mozer et al., 2020]. Figure 5 of Mozer et al [2021] shows a continuous 30-minute segment of a TIAW that actually lasted for hours, as can be seen from the spectra of Figure 1 in that paper. Such waves appear when the ratio of the core electron to ion temperature, Te/Ti, is greater than one, as seen in Table 1 of this paper and Figures 1-4 of Mozer et al [2022a]. In such an environment, Landau damping is diminished and the waves can survive for longer times. This may explain why TIAW are observed for long times.

**IIh The high frequency wave in TIAW is narrow band. Such narrow band waves are not found in ordinary ion-acoustic waves. Formation of this wave by coupling with the low frequency wave can explain why the high frequency wave is narrow band.**

The following discussion suggests that the source of the monochromatic high frequency TIAW is a resonant mechanism that creates only a single monochromatic electrostatic wave which, with time, produces electrostatic waves at higher harmonic frequencies, as are observed.

Depending on the formation mechanism, an ion-acoustic wave may initially contain any number of discrete frequencies, n, where n≥1. The value n=1 is possible although previous observations in the solar wind [Kurth et al., 1979] and the Earth's bow shock [Hull et al., 2006; Vasko et al., 2022] showed that n is usually large because ion-acoustic waves typically have spectral widths comparable with the wave central frequency. Because the high frequency TIAW generally lasts for a short time and at a specific phase of the low frequency wave, it appears to be the result of a process that occurs because the plasma is unstable at a specific phase of the low frequency wave and n=1.

After formation, ion-acoustic waves steepen to create harmonics of the initial wave due to fluid nonlinearities [e.g., Davidson, 1972]. The steepening time scale can be estimated as $\lambda/2\pi\,\delta u$, where $\delta u \sim eE/2\pi f_h\, m_i$ is the amplitude of the ion bulk velocity fluctuations due to the electrostatic wave, and $\lambda$ and $f_h$ are the wavelength and wave frequency in the plasma rest frame. Using the observed amplitudes and assuming a reasonable phase speed of 200 km/sec for the ion-acoustic waves in the plasma frame, the steepening time scale of the TIAW high-frequency waves is a few seconds. A second nonlinear mechanism that produces harmonics of an initially monochromatic ion-acoustic wave, is ion trapping by the electrostatic wave [e.g. Davidson, 1972] and this process operates on a time scale of $\lambda/2\pi\delta u$, where $\delta u \sim (e\delta\varphi/m_i)^{1/2}$ is the speed of the trapped ions, which is a few seconds as well. This widening produce waves at harmonics of the initial frequencies.

Figure 6 presents a rare example in which the high frequency wave of (6d) is continuous in time over the duration of the low frequency wave of (6e), so formation of harmonics should be observed. In panels (6a) and (6b), the 0-1000 Hz spectra of the electric field and density fluctuations show the fundamental wave at about 225 Hz and the first harmonic at about 450 Hz. Also, the next higher harmonic is barely visible in the electric field spectrum. In panel (6c) the same harmonic generation of the low frequency wave is observed in the 0-10 Hz spectral plot. Note that the steepening creates harmonics of the narrow band wave and does not broaden the spectrum of this wave.

**IIi The low frequency wave in TIAW is a narrow band ion-acoustic wave.**

Like the high frequency wave in TIAW, the low frequency wave is also nearly monochromatic. Figure (5a) shows that this is true for EX, and Figure (6c), which displays the power spectrum from 0 to 10 Hz, illustrates that the fundamental low frequency signal was a monotone at about 4 Hz.

A theory that might explain the generation of the low frequency ion-acoustic wave in the solar radial range of the present observations is based on the conservation of energy and the first invariant during the solar wind expansion [Kellogg, 2022]. This creates a hole in the ion distribution near zero velocity, which results in an instability that produces a low frequency electrostatic wave.

A different low frequency electrostatic wave mode may be generated if the plasma contains a mixture of electrons, protons and charged dust particles [Malaspina et al [2020]; Segwal and Sharma [2018]]. However, to create a dust wave at a frequency ~1 Hz requires a dust density many orders-of-magnitude greater than that observed, so the low frequency TIAW cannot be a dust wave.

> **IIj The <0.1 Hz electric field measured at times of the observed TIAW was in agreement with -vxB, the motional electric field, which suggests that the instrument functioned nominally.**

Figure 7 presents four hours of data that illustrate this fact. Panel (7a) of this figure gives the power spectrum of Ex which, on careful viewing, is seen to be composed of pulses of energy corresponding to the bursts of the high frequency component of the phase locked wave. Thus, triggered ion-acoustic waves were present through most of the interval. Panels (7b) and (7c) present the two components of the electric field and the components of –**vxB**, the motional electric field, all of which are band pass filtered below 0.1 Hz. The agreement between the electric field measurements and –**vxB** shows that the very low frequency electric fields were well-measured. These very low frequency electric fields were produced by least-squares fitting the electric field to –**vxB**. The two least-squares coefficients produced in this way are the effective antenna length of panel (7d) and the angular rotation of the X-Y plane in panel (7e) [Mozer et al, 2020a]. The reasonableness of these quantities as well as the excellent agreement between **E** and **–vxB** show that the very low frequency electric field was well-measured, which suggests that there is no concern about the instrument operation during this time.

> **IIk The TIAW were not affected by variations of the plasma density or temperature, the solar wind speed, magnetic field orientation, etc., indicating that the measurement was not made in a wake or other effect that produced a spurious field.**

To test the possibility that plasma parameter variations can affect the TIAW, as was the case in Malaspina et al [2022], Figure 8 gives the solar wind speed in panel (8c), the angle of the magnetic field from the Z-direction in panel (8d), the plasma density in panel (8e), and the ion temperature in panel (8f) during the four interval of TIAW illustrated in Figure 7. During this interval, the solar wind speed varied by 50%, the angle of the magnetic field relative to the Z-direction varied by 90 degrees, and there were important changes in the plasma density and temperature. These variations affected the magnitudes of E and -vxB but they did not affect the quality of the fit of these two parameters, which provides evidence that wake or other near-spacecraft effects did not infuence the electric field measurements.

> **IIl The single ended potentials of the electric field measurement show that it performed normally.**

Further evidence that the measured fields are physical comes from examination of the four antenna potentials, typical examples of which are presented in Figure 9. The four antenna voltages, filtered from 1-100 Hz, are presented in panels (9a), (9b), (9c), and (9d), while the same data, filtered above 100 Hz, are in the bottom four panels. At both the low and high frequencies, all four antennas produced good electric potentials with no indication that one or more of the potentials was perturbed by a wake or any other non-physical effect. Because this behaviour existed during all events, it represents strong evidence that the antennas performed nominally and that the triggered ion-acoustic waves are a real physical phenomenon.

**IIm The mechanism for the coupling of the two electrostatic waves in TIAW is not understood, although similar types of coupling are known.**

The mechanism resulting in phase-locking between the high- and low-frequency electrostatic waves is not understood. Other observations have demonstrated that phase correlation between electrostatic spikes and whistler waves can occur due to the steepening process of the whistler wave electrostatic component under the resonance condition between electron-acoustic and whistler waves [Agapitov et al., 2018; Vasko et al., 2018; An et al., 2019]. The coupling between Langmuir and ion-acoustic waves has also been shown theoretically [Nishikawa et al, 1974].

Recent simulations showed that electrostatic fluctuations can be observed at a specific phase of an oblique whistler wave, since the drift between cold and hot electrons associated with the whistler wave results in a high-frequency electrostatic instability [Roytershteyn and Delzanno, 2021]. Following this idea, one may hypothesize that the phase-locking observed in triggered ion-acoustic waves is due to a current-driven instability producing high-frequency electrostatic waves at the specific phase of the low-frequency electrostatic wave. In this case, the phase-locking is observed because the high-frequency waves are at the phase where the plasma is unstable. The viability of this mechanism for triggered ion-acoustic waves deserves a separate study, while here we only point out that the ion-electron drift velocity associated with the low-frequency waves is negligible compared to the ion-acoustic speed and, thus, not likely to be unstable.

III SUMMARY

In summary, the above results show that the electric field instrument functioned normally (IIj and Il), that it measured two electrostatic waves (IIb) that had the nearly same phase velocity (III) and that satisfied the requirements of electrostatic waves because the electric field and density fluctuations were 90 degrees out of phase (IIf), the electric field was parallel to the k-vector (IIc), and the phase velocity was consistent with that expected for an ion-acoustic wave (IId). These linked waves were found on most orbits between 20 and 25 solar radii in regions where $T_e/T_i>1$ (due to a low ion temperature) such that ion Landau damping was small and they might survive for hours (IIg). The electric field and density fluctuations in the high frequency wave were nearly monochromatic (IIh). The low frequency wave was also narrow band with harmonics and it was an ion-acoustic wave and not a dust mode (IIi).

The triggered ion-acoustic waves did not depend strongly on plasma parameters such as the ion temperature, the density, the magnetic field, or the solar wind speed (IIk). They were present during seven orbits in the region where electrons were heated (IIa) and absent when there was little or no electron heating. The potential of the low frequency wave was sufficient for it to provide the energy that did the heating (IIa). Why or how the high and low frequency components of the triggered ion-acoustic wave were coupled is not understood but similar phenomena have been found between other wave pairs (IIm). Thus, the evidence that the triggered ion-acoustic waves are a real physical phenomenon that is important to the solar wind physics is very strong, although a non-physical source cannot be completely ruled out. That it cannot be ruled out as an artifact is the case for almost any physical observation made in space.

IV ACKNOWLEDGEMENTS

This work was supported by NASA contracts NNN06AA01C and 80NSSC21K0581. The authors acknowledge the extraordinary contributions of the Parker Solar Probe spacecraft engineering team at the Applied Physics Laboratory at Johns Hopkins University. The FIELDS experiment on the Parker Solar Probe was designed and developed under NASA contract NNN06AA01C. Our sincere thanks to P. Harvey, K. Goetz, and M. Pulupa for managing the spacecraft commanding, data processing, and data analysis, which has become a heavy load thanks to the complexity of the instruments and the orbit. We also acknowledge the SWEAP team for providing plasma data. The work of I.V. was supported by National Science Foundation grant No. 2026680. The work of J.V. was supported by NASA PSP-GI 80NSSC23K0208.

TABLE 1

PLASMA PARAMETERS AT 01:19:15 ON 06/01/2021

| Magnetic field | 200 | nT |
|---|---|---|
| Density | 1100 | $cm^{-3}$ |
| Ion temperature | 20 | eV |
| Core electron temperature | 54 | eV |
| Solar wind speed | 190 | km/sec |
| Solar wind vX | -85 | km/sec |
| Solar wind vY | -4 | km/sec |
| Solar wind vZ | -170 | km/sec |
| Ion beta | 0.2 | |
| Proton Debye length | 0.8 | m |
| Alfven speed | 100 | Km/sec |
| Electron gyrofrequency | 6000 | Hz |

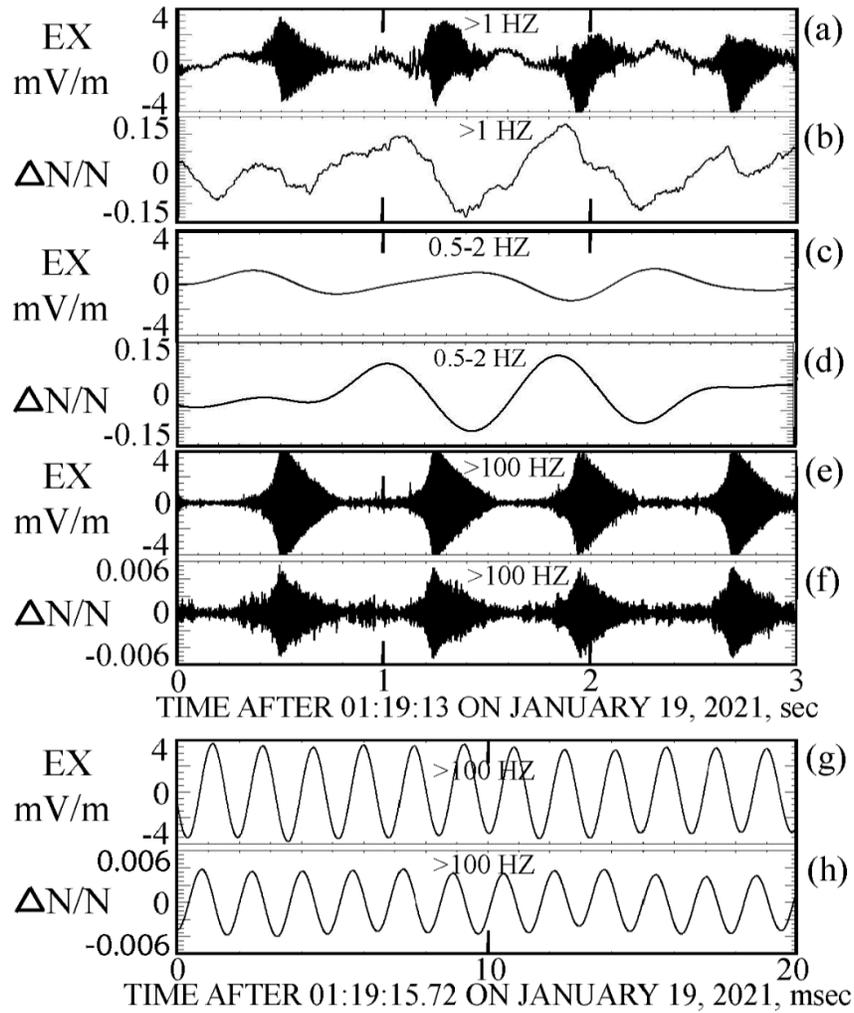

Figure 1. Four pairs of measurements of EX and density fluctuations. The pair (1a), (1b) are high pass filtered at 1 Hz to display the phase locked low and high frequency signatures of the triggered ion-acoustic wave. Panels (1c) and (1d) are bandpass filtered at 0.5-2 Hz to display the low frequency components of the wave. The pair (1e), (1f) are high pass filtered at 100 Hz to show the high frequency components. The pair (1g), (1h) provide >100 Hz filtered data over a short time interval to illustrate the pure sine wave nature of the high frequency component.

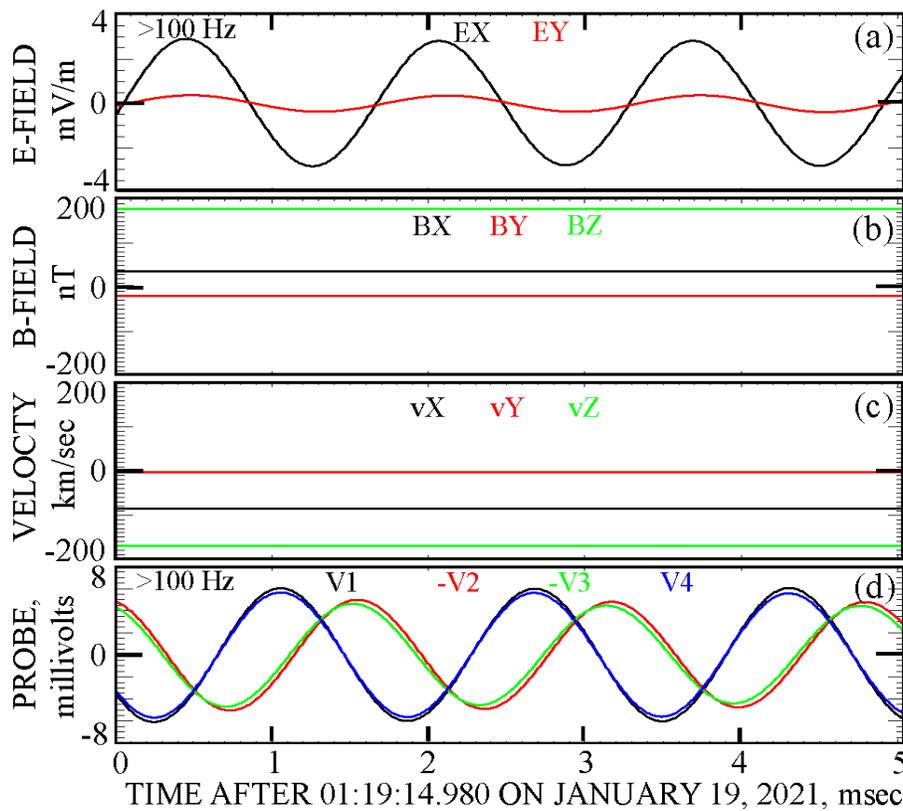

Figure 2. The >100 Hz EX and EY of the TIAW in panel (2a), the 200 nT background magnetic field vectors in the X-Y-Z frame in panel (2b), the 200 km/sec solar wind velocity in panel (2c), and the timing of the signals received by the four antennas for the event of Figure 1 in panel (2d).

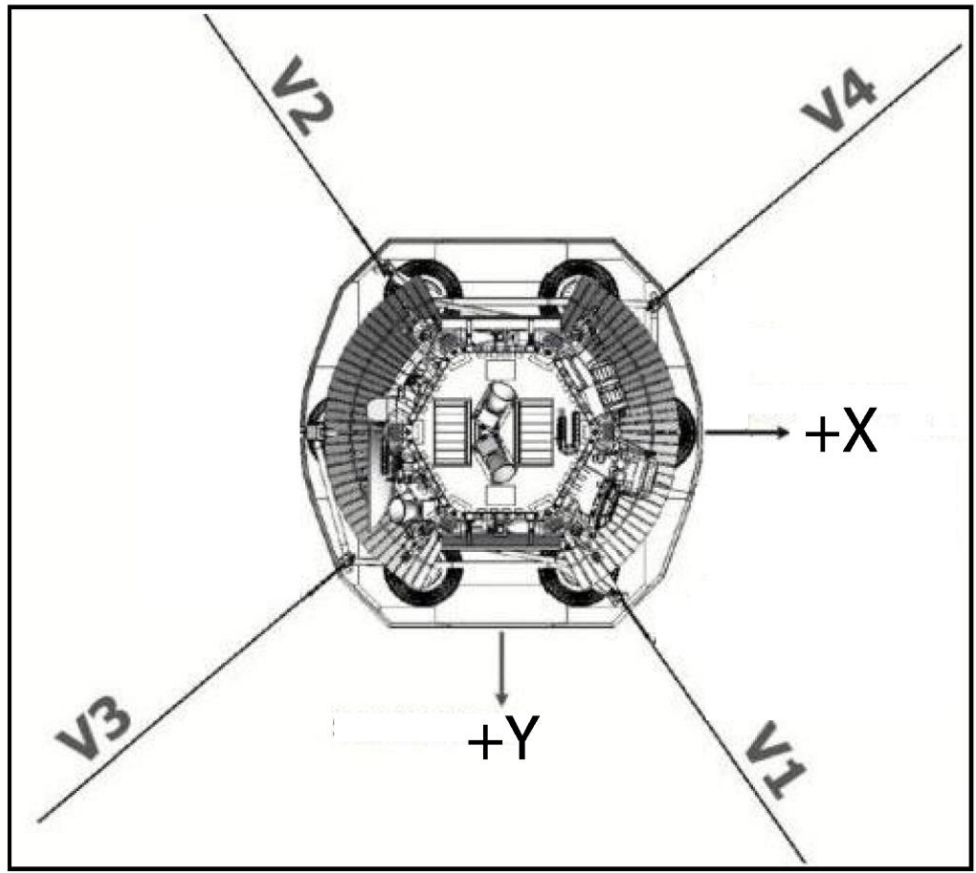

Figure 3. Orientation of the electric field antennas with respect to the X and Y axes.

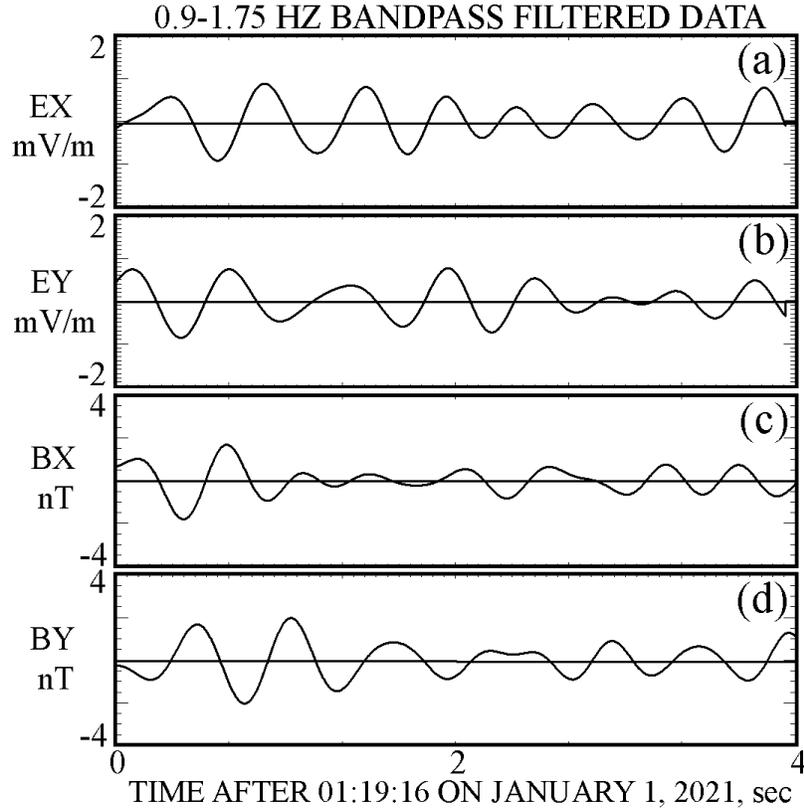

Figure 4. Two components of the electric field in panels (4a) and (4b) and the magnetic field in panels (4c) and (4d). Because EX is nearly a pure sine wave, the electrostatic E-field in this component greatly exceeded the electromagnetic E-field. By contrast EY correlated with BX to suggest that the electrostatic E-field was smaller than the electromagnetic E-field in this component.

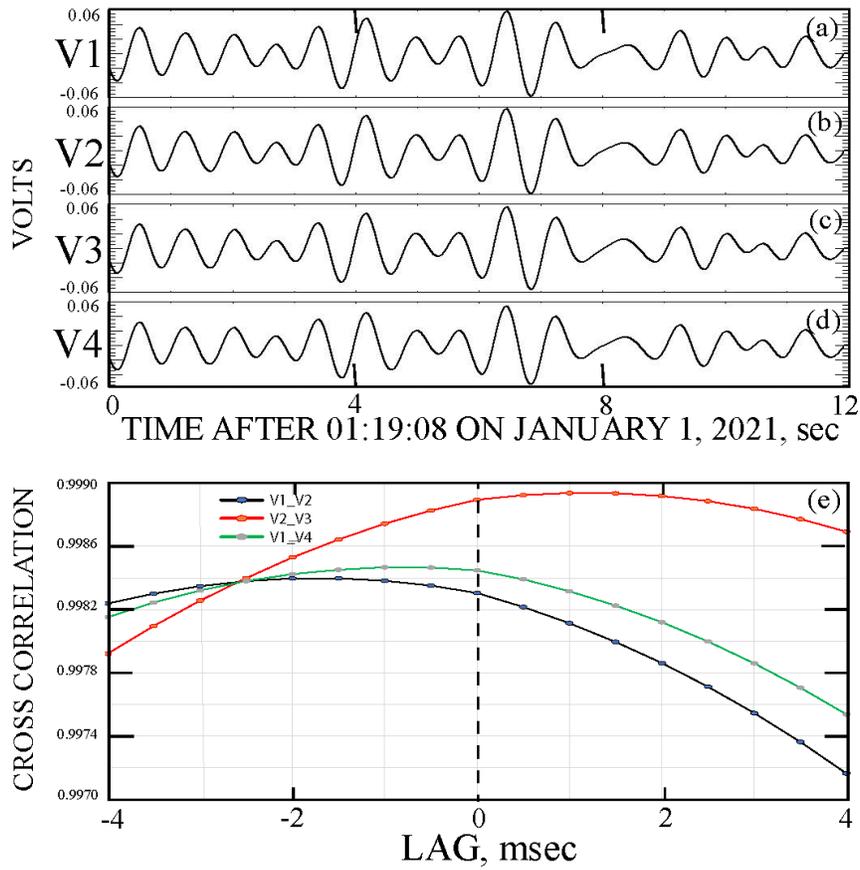

Figure 5. The four antenna potentials in panels (5a), (5b), (5c), and (5d), covering the time interval of Figures 1 and 2, and band pass filtered between 0.95 and 1.75 Hz. The phase lags between antenna pairs in panel (5e) give the times required for the low frequency wave to travel between antenna pairs and lead to the conclusion that, in the X-Y plasma frame, the low frequency wave's phase velocity has the same magnitude and direction as does the high frequency wave.

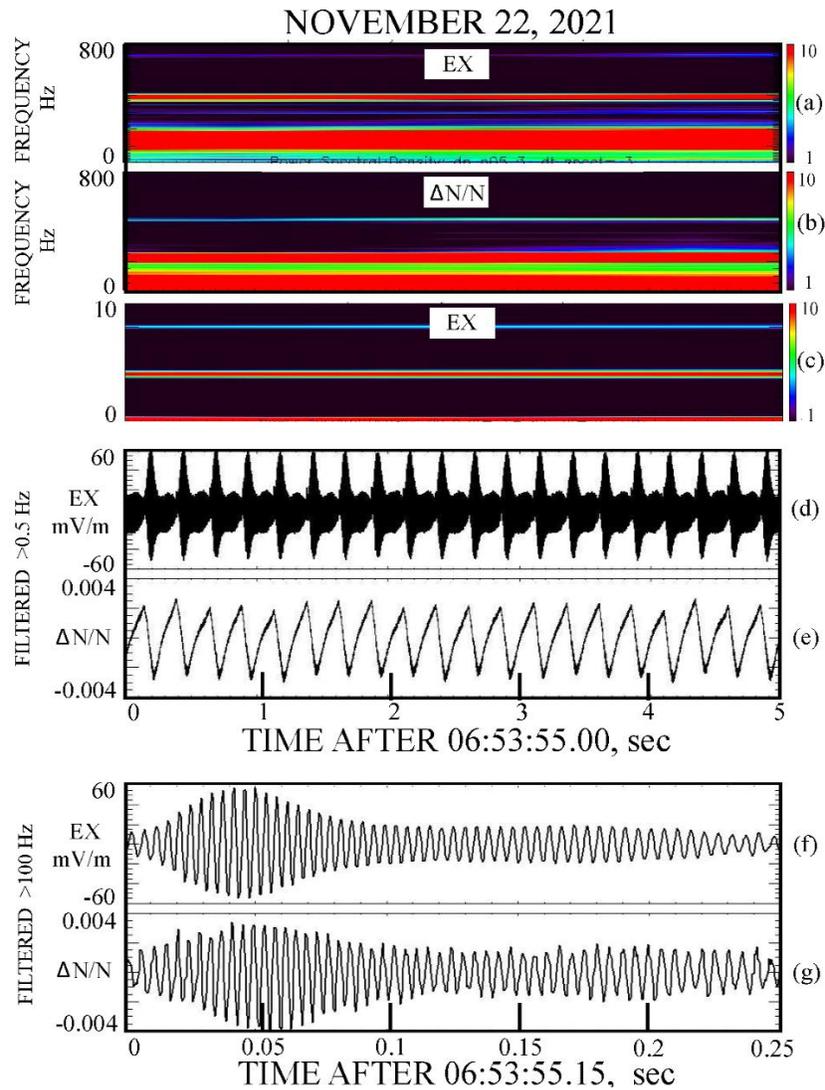

Figure 6. Illustration of a triggered ion-acoustic wave whose high frequency component appeared continuously during the low frequency component, as seen in panels (6c) and (6d), which present the >0.5 Hz filtered EX electric field and the density fluctuations. Panels (6e) and (6f) present 0.25 second segments of the >100 Hz fields and density fluctuations in order to illustrate the pure sine wave nature of the signals that persist throughout the low frequency signal. Because these high frequency signals are present continuously, it is expected that they should steepen to produce harmonics of the fundamental frequency. Panels (6a) and (6b) present the 0-1000 Hz spectra of the electric field and density fluctuations that show both the presence of the fundamental signals at about 225 Hz, and also the first harmonics at about 450 Hz. In the electric field spectrum, there is also a weak signature at the next harmonic. Panel (6c) presents the 0-10 Hz spectrum of the electric field which shows the narrow band low frequency wave at 4 Hz and its first harmonic at 8 Hz.

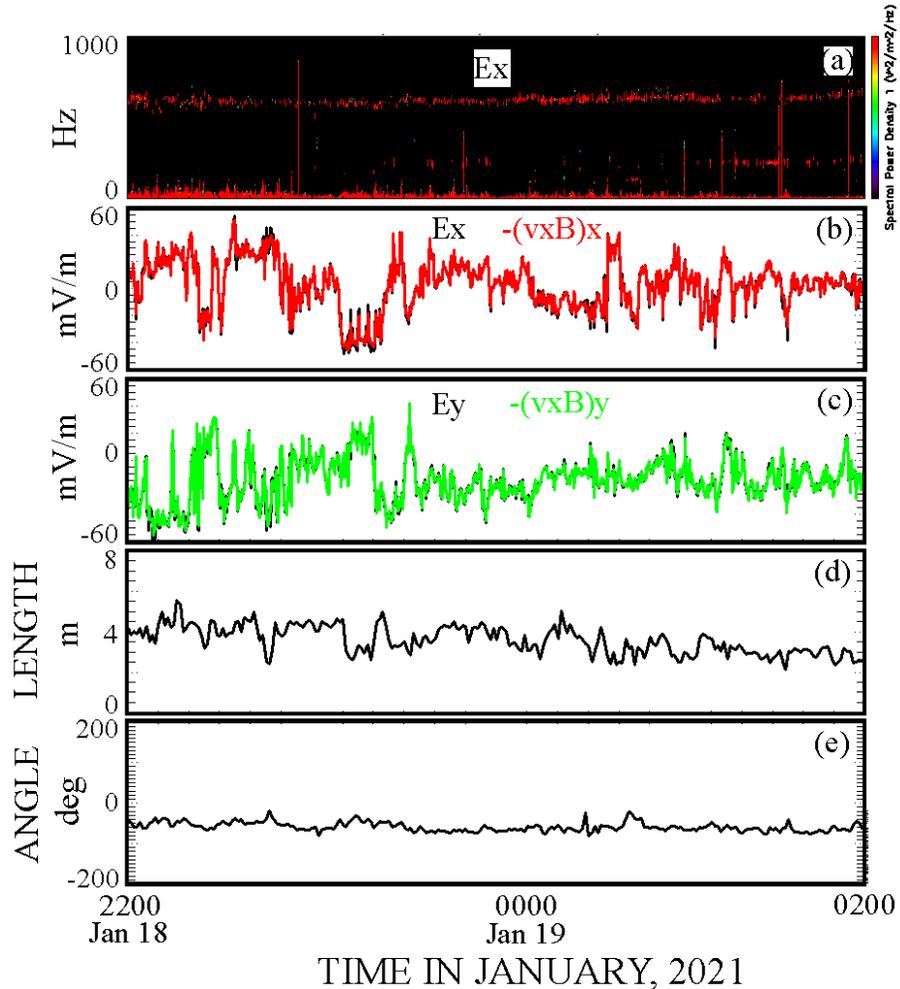

Figure 7. A four hour interval illlustrating the accuracy of the <0.1 Hz electric field measureents. Panel (7a) gives the electric field spectrum which, on close examination, consists of a series of red dots showing that the 600 Hz pure sine wave occurred in triggered bursts through the interval. Panel (7b) gives the dc and low frequency X-component of the electric field and the X-component of **–vxB**. Panel (7c) provides the same data for the Y-compnents. The agreement between **E** and –**vxB** shows that the instrument functioned normally during this interval. The electric field is obtained from a least-squares fit described in Mozer et al [2020a], and the two least-squares coefficients are shown in panels (7d) and (7e).

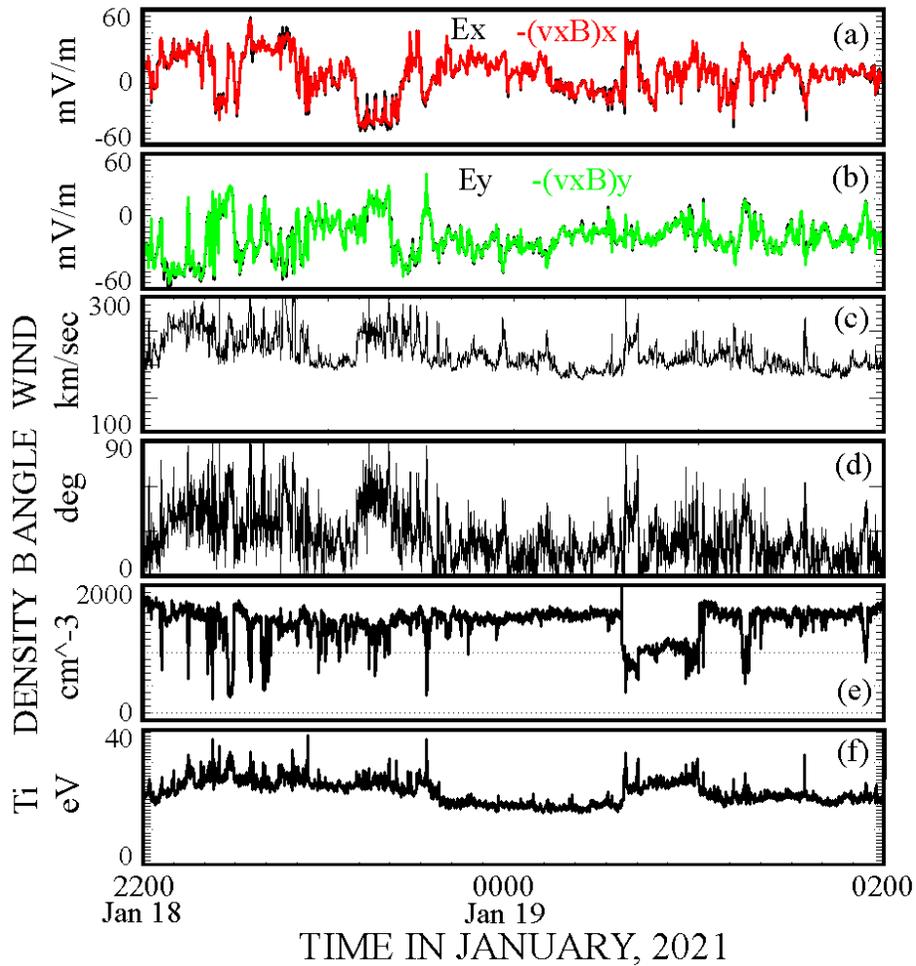

Figure 8. The solar wind speed (8c), the angle of the magnetic field with respect to the Z-direction (8d), the plasma density (8e), and the ion temperature (8f). During the four hour interval of interest, the magnetic field rotated through as much as 90 degrees, the solar wind speed varied by 50%, and the temperature and density had significant variations, all of which occurred during electric field measurements that agreed with -vXB. This result suggests that the electric field data were not perturbed by wake or similar effects.

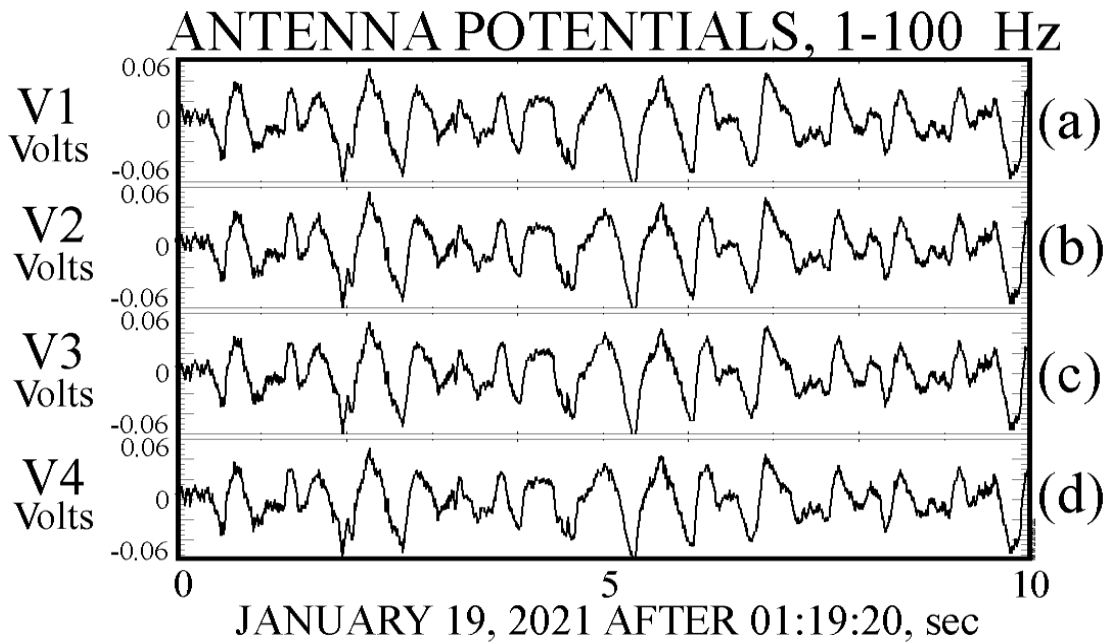

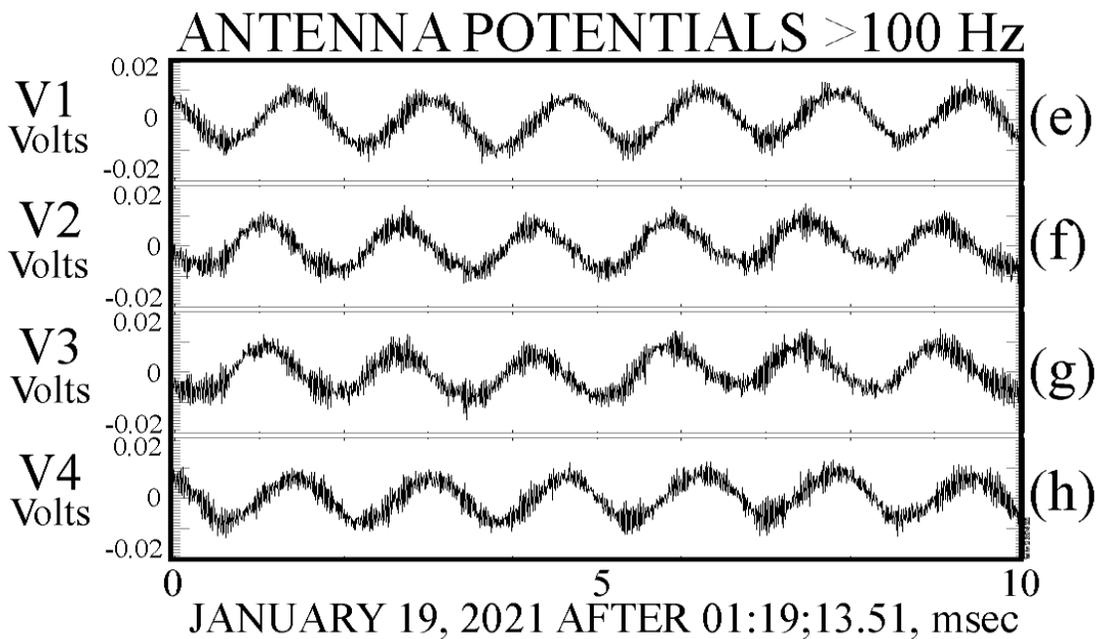

Figure 9. Typical four antenna potentials (V1, V2, V3, and V4) at the two frequencies of the triggered ion-acoustic waves. In both cases, all of the antennas operated normally with no indication of one or more of them being affected by a wake or other non-physical process. This provides strong evidence that the electric field instrument performed normally and that the observed waves were physical and not an artifact resulting from a wake or detector malfunction.
20...